\documentclass[aps,asmsymb,twocolumn,showpacs,amsmath,prb]{revtex4-1}

\usepackage{graphicx}
\usepackage{subfigure}
\usepackage{epsfig}
\usepackage{dcolumn}
\usepackage{bm}
\usepackage{color}
\usepackage{xcolor}
\usepackage{bm}
\usepackage{hyperref}
\usepackage{amsmath, amssymb, dsfont}

\def\be{\begin{equation}}       \def\ee{\end{equation}}
\def\bea{\begin{eqnarray}}      \def\eea{\end{eqnarray}}
\def\ba{\begin{array}}
\def\ea{\end{array}}
\def\bnum{\begin{enumerate} }
\def\enum{\end{enumerate}}

\def\=>{\Rightarrow}
\def\>{\rightarrow}

\def\eye2{Fathbb{I}}
\def\bk{{\bf k}}
\def\bh{{\bf h}}
\def\bJ{{\bf J}}

\def\br{{\bf r}}
\def\si{{\sigma}}

\renewcommand{\>}{\rangle}

\newcommand{\al}{\alpha}

\begin{document}

\title{Imbert-Fedorov shift in Weyl semimetals: Dependence on monopole charge and intervalley scattering}
\author{Luyang Wang$^{1,2}$ and Shao-Kai Jian$^{1}$}
\affiliation{$^1$Institute for Advanced Study, Tsinghua University, Beijing 100084, China\\
$^2$State Key Laboratory of Optoelectronic Materials and Technologies, School of Physics, Sun Yat-Sen University, Guangzhou 510275, China}

\begin{abstract}
The Imbert-Fedorov (IF) shift in optics describes the transverse shift of light beams at the reflection interface. Recently, the IF shift of Weyl fermions at the interface between Weyl semimetals (WSMs) with single monopole charge has been studied. Here, we study the IF shift at the interface between two WSMs, each of which is carrying an arbitrary integer monopole charge. We find a general relation between the monopole charges of the two WSMs and the IF shift. In particular, the IF shift is proportional to the monopole charge if both WSMs have the same one. Our results can be used to infer the topology of the materials by experimentally measuring their IF shift. Furthermore, we consider the possibility that the Weyl fermions are scattered to other Weyl cones during the reflection, which results in qualitatively different behavior of the IF shift. While we use a quantum mechanical approach to solve the problem, semiclassical equations of motion and the conservation of total angular momentum can help us intuitively interpret our results in special cases.
\end{abstract}
\date{\today}
\maketitle	

\section{introduction}
Spin-orbit coupling (SOC) plays an increasingly important role in modern condensed matter physics. It is essential in many topological phenomena, such as the spin Hall effect\cite{sinova-2015}, the quantum spin Hall effect\cite{maciejko-2011} and topological insulators\cite{hasan-2010,qi-2011}. In recent years, SOC has also attracted much interest in optics\cite{bliokh-2015}. Due to the intrinsic SOC of light, it has been realized that the spin Hall effect naturally occurs in optical systems when there is a gradient in the refractive index, resembling an electric field in electronic systems. As a result, at the interface between two media where the refractive index varies, when transmitted or reflected, a light beam has a transverse shift with the direction depending on the chirality (spin) of the photons\cite{onoda-2004,bliokh-2006}. The transverse shift is named the Imbert-Fedorov (IF) shift after its discoverers\cite{fedorov-1955,imbert-1972}, and has been verified by experiments\cite{hosten-2008,yin-2013}.

In recent works\cite{xie-2015,yang-2015b}, the authors have found that the IF shift also occurs in Weyl semimetals (WSMs). WSMs are recently discovered materials\cite{lv-2015a,xu-2015a,yang-2015a,lv-2015b,xu-2015b}. They host Weyl points, which behave as monopoles of Berry flux. Around the Weyl points, the states are described by the Weyl equation\cite{weyl-1929} and have a linear dispersion\cite{wan-2011,volovik-2009,xu-2011,burkov-2011,yang-2011,halasz-2012,zhang-2014,liu-2014,weng-2015,huang-2015,hirayama-2015}. The IF shift in WSMs is due to the intrinsic coupling between the pseudospin and the orbital degree of freedom. It has been shown that the IF shift depends on the monopole charge (chirality) of the Weyl point which is $\pm1$, similar to that in optical systems.

The IF shift in both optical systems and WSMs has been interpreted semiclassically\cite{onoda-2004,yang-2015b}. The semiclassical equations of motion (EOM) govern the trajectory of wave packets, and dictate that the IF shift of a wave packet is due to its anomalous velocity\cite{xiao-2010}, and hence is an integral of the Berry curvature. If the band structure and hence the Berry curvature vary slowly, then it is an intuitive way to calculate the IF shift. However, if the Berry curvature has an abrupt change at an interface, difficulties arise. Moreover, near the Weyl point where the gap closes, non-Abelian treatment is needed\cite{yang-2015b}. As such, processes like Klein tunneling are not straightforward to account for. Different from the quantum mechanical treatment, the trajectory of a wave packet is fixed, which ignores the possibility that the wave packet may split. Another interpretation of the IF shift is from the conservation of total angular momentum (TAM)\cite{onoda-2004,yang-2015b}, which only works if the two media have the same monopole charge, and the composite system has a rotational symmetry. Therefore, there is a limit in these two methods. In WSMs, the quantum mechanical approach has been applied to the calculation of the IF shift\cite{xie-2015}. It works best if an abrupt interface exists between two WSMs, regardless of the symmetry of the system. Each of the three approaches has its own advantages and drawbacks.

Multi-WSMs that host Weyl points with monopole charge $\pm2$ and $\pm3$ have been discovered\cite{fang-2012}, which are protected by point group symmetries. Namely, $C_4$ and $C_6$ symmetries can protect double-Weyl points while $C_6$ symmetry can protect triple-Weyl points. Therefore, it is natural to extend previous works on the IF shift in single-WSMs to multiple WSMs. In this work, we study the IF shift between WSMs with monopoles of arbitrary integer charges. We apply the quantum mechanical approach to study the problem, and then use the semiclassical approach and the conservation of TAM to confirm and interpret the results in special cases. In addition, we consider the possibility that the Weyl fermions are reflected to another Weyl cone with opposite chirality, in which case the IF shift vanishes under certain symmetry conditions.

\section{IF shift at the interface between WSMs with arbitrary monopole charges}
\begin{figure}
  \centering
  \includegraphics[width=.4\textwidth]{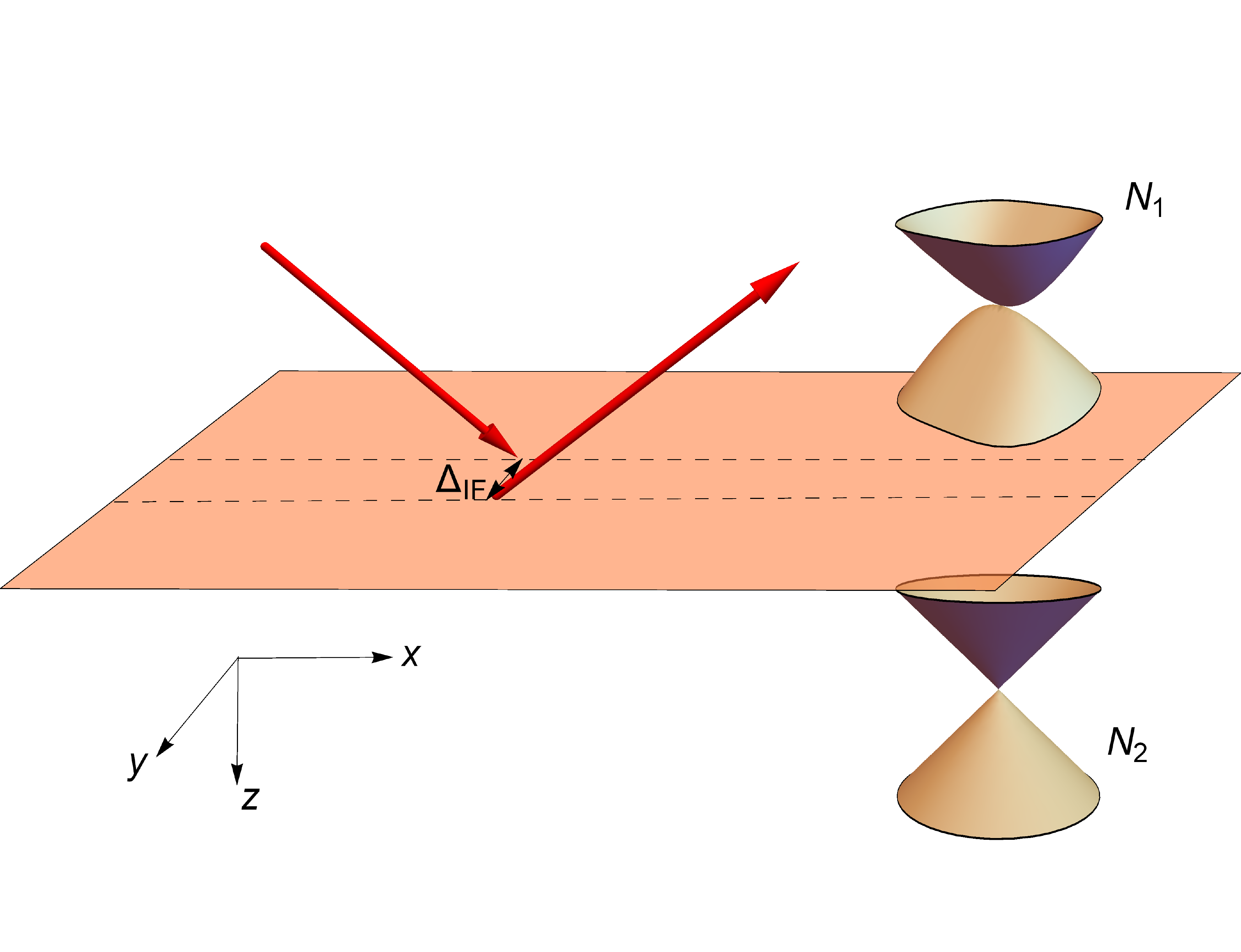}
  \caption{A beam of Weyl fermions is incident from a WSM at $z<0$, with monopole charge $N_1$, to another at $z>0$, with monopole charge $N_2$, and is totally reflected at the interface $z=0$. A step potential is assumed between the two media. During the reflection, the IF shift occurs. }\label{fig:refl}
\end{figure}
The Hamiltonian for a Weyl cone with a positive monopole charge $N$ can be written as
\be
H_0(\bk)=\left(\begin{array}{cc}k_z&(k_x-ik_y)^N\\(k_x+ik_y)^N&-k_z\end{array}\right),\label{eq:ham1}
\ee
where $\bk=(k_x,k_y,k_z)$ and $H_0^*$ carries the opposite charge $-N$. We have chosen the $z$-axis as a high symmetry axis. We set the velocities to 1 hereafter for simplicity. (Actually, the IF shift depends on the velocities\cite{xie-2015,yang-2015b}; we leave the calculation of such dependence to future work.) Note that the generalization of a single-Weyl point with monopole charge 1 to a multi-Weyl point with monopole charge $N$ is strongly reminiscent of the counterpart in two dimensions –- the generalization of a gapless Dirac point with winding number 1 to the case with winding number $N$\cite{zhang-2011}. The eigenenergies are $E=\pm\sqrt{k_\parallel^{2N}+k_z^2}$, associated with the spinor wavefunction
\begin{eqnarray}
  \psi(\br) &=& \frac{e^{i(k_xx+k_yy+k_zz)}}{\sqrt{1+\eta^2}}\left(\begin{array}{c}
                                                                        e^{-iN\al} \\
                                                                        \eta
                                                                      \end{array}\right)
\end{eqnarray}
where ${\bf r}=(x,y,z)$, $k_\parallel=\sqrt{k_x^2+k_y^2}$, $\al=\tan^{-1}{(k_y/k_x)}$ and $\eta=\sqrt{\frac{E-k_z}{E+k_z}}$.

We consider the total reflection that occurs at the interface between two WSMs. Assume a beam of Weyl fermions is incident from a WSM with monopole charge $N_1$ to that with monopole charge $N_2$, and the interface is at $z=0$, as shown in Fig.\ref{fig:refl}. A step potential is assumed at the interface, with the form $V=0$ for $z<0$ and $V=V_0>0$ for $z>0$. During the process of reflection, $k_x$ and $k_y$ are conserved, while $k_z$ is not. Using the continuity of the wavefunction at $z=0$ and solving for the reflective coefficient $r$, we get $r=e^{i\phi_r}$
with the phase (see Appendix A for details)
\begin{eqnarray}
  \phi_r &=& 2\tan^{-1}\frac{\eta\sin(\phi_\xi-(N_1-N_2)\al)}{1-\eta\cos(\phi_\xi-(N_1-N_2)\al)}\nonumber\\
  &+&\phi_\xi-(N_1-N_2)\alpha
\end{eqnarray}
where $\phi_\xi$=$\arg(E-V_0-i\kappa)$ with $\kappa=\sqrt{k_\parallel^{2N_2}-(E-V_0)^2}$. The total reflection only occurs when $\kappa>0$ since the wave function is proportional to $e^{-\kappa z}$ at $z>0$. We assume the incident beam is Gaussian in the $y$-direction, and the central ray lies in the $xz$ plane. Then the $k$-space distribution is also Gaussian, with the center at $\bar k_y=0$. In real space, the center of the incident beam is $N_1\partial_{k_y}\al/(1+\eta^2)|_{k_y=0}$, and the center of the reflected beam is $[(N_1\eta^2\partial_{k_y}\al)/(1+\eta^2)-\partial_{k_y}\phi_r]|_{k_y=0}$, so the IF shift is their difference. Bringing all the variables in, we find the IF shift as follows:
\begin{eqnarray}
  \Delta_{IF} &=&-\frac{N_1k_z}{Ek_x}+\frac{(N_1-N_2)k_z}{Ek_x+(V_0-E)k_x^{N_1-N_2+1}}.\label{eq:DIF1}
\end{eqnarray}

Two special cases are noticed. First, if the two WSMs have the same monopole charge, i.e. $N_1=N_2\equiv N$, then $\Delta_{IF}=-N/(E\tan\theta)$, where $\theta=\tan^{-1}k_x/k_z$. Note that $\theta$ is different from the incident angle, since the velocity is not proportional to the momentum in general. (Previous studies on single-Weyl points correspond to $N=\pm1$\cite{xie-2015,yang-2015b} here.) 
The IF shift changes sign once the chirality changes, which is termed as the chirality-dependent Hall effect. Here, we extend it to a general monopole charge, and find the IF shift is proportional to the monopole charge. This is like the case in the quantum Hall effect, where the quantized Hall conductance is proportional to the Chern number\cite{thouless-1982}. Second, if the energy of the incident Weyl fermions can be adjusted to $E=V_0$, then $\Delta_{IF}=-N_2/(E\tan\theta)$. Then one can detect the monopole charge of the WSM to which the Weyl fermions are incident.

One disadvantage of the quantum mechanical approach is that we need to assume a specific form of the wave packet to calculate the IF shift. In addition, we have assumed a sharp interface between the two WSMs. However, the Weyl Hamiltonian is actually a low energy, long wavelength Hamiltonian, so the matching wavefunction method only works for the case in which the interface is smooth in the atomic scale. Therefore, we need to confirm that the result obtained above is valid. To this end, we apply the semiclassical approach to calculate the IF shift where the interface is smooth.

\section{Semiclassical approach}
Analogous to the motion of electrons in condensed matter systems, the motion of optical wave packets can be studied by semiclassical EOM\cite{onoda-2004,onoda-2006}, given the condition that the modulation to the band structure is weak and slowly varying. The semiclassical approach gives a simple interpretation to the IF shift: it is the shift due to the anomalous velocity, which appears if the Berry curvature is nonvanishing and an effective electric field exists\cite{xiao-2010}. This approach is used to calculate the IF shift in WSMs in Ref.\cite{yang-2015b}, in the case where the band structure and the Berry curvature are slowly varying. Now we show that the IF shift between two WSMs with the same monopole charge $N$ can be calculated in this way, which further confirms the results obtained above.

The incident momentum of a wave packet is $(k_x,0,k_z)$, and the reflected momentum is $(k_x,0,-k_z)$. The IF shift is the integral of the anomalous velocity\cite{onoda-2004},
\begin{eqnarray}
\Delta_{IF}=\int dt[\dot{\bk}\times{\bf\Omega}(\bk)]_y=\int_{k_z}^{-k_z}dk_z \Omega_x(k_x,0,k_z),\label{eq:DIF}
\end{eqnarray}
where $\dot {\bf k}$ denotes the time derivative of ${\bf k}$, ${\bf \Omega(\bk)}$ is the Berry curvature of the band in which the wave packet resides and $[...]_i$ means the $i$-th component. We see the difference between the Chern number and the IF shift: the former is the integral of the Berry curvature over the planar Brillouin zone, while the latter is the integral of the Berry curvature over a line segment in the Brillouin zone.

The Hamiltonian Eq.(\ref{eq:ham1}) can be written as $H_0(\bk)=\bh(\bk)\cdot {\bm \sigma}$, where components of ${\bm \sigma}$ are Pauli matrices and components of $\bh$ are the coefficients of each Pauli matrix, respectively. Then the $x$-component of the Berry curvature is ${\Omega}_x(\bk)=\bh\cdot(\partial_{k_y}\bh\times\partial_{k_z}\bh)/(2h^3)$, where $h=\sqrt{ \sum_{i=1}^3 h_i^2}$. To obtain $\Omega_x(k_x,0,k_z)$, we expand $\bh(\bk)$ to the first order of $k_y$, $\bh(\bk)=(k_x^{N}, Nk_x^{N-1}k_y,k_z)+O(k_y^2)$, and then find $\Omega_x(k_x,0,k_z)=Nk_x^{2N-1}/[2(k_x^{2N}+k_z^2)^{\frac{3}{2}}]$. Plugging it into Eq.(\ref{eq:DIF}), we have $\Delta_{IF}=-N/(E\tan\theta)$, which agrees with the quantum mechanical result.

An immediate question is whether we can apply the semiclassical approach to the case where the two WSMs carry different monopole charges. A necessary condition to apply the semiclassical approach is one can interpolate the two Hamiltonians smoothly, such that the band structure and the Berry curvature vary slowly. However, since the two Hamiltonians have different topologies, i.e. their monopole charges differ by an integer, they cannot be connected smoothly. Therefore, the semiclassical approach cannot be applied to this case.

\section{Conservation of TAM}
The IF shift in optics has also been understood as the result of the conservation of the total angular momentum of individual photons\cite{onoda-2004}. In WSMs with $N=1$, it has been shown that the IF shift can be interpreted as the result of the conservation of the generalized total angular momentum, which is the sum of the orbital angular momentum and the $1/2$ pseudospin of Weyl fermions\cite{yang-2015b}. Here we generalize further this idea.

If both WSMs are described by $H_0+V$ where $H_0$ is given in Eq.(\ref{eq:ham1}) and $V$ is a step potential, i.e. both have the same monopole charge $N$, there is a continuous rotational symmetry around the $z$-axis, corresponding to the conservation of the $z$-component of the total angular momentum $\bJ$. One can show $[J_z,H]=0$, where $J_z=L_z+S_z$, with $L_z=xk_y-yk_x=-i\partial_\al$ and $S_z=\frac{N}{2}{\bf\si}_z$. 

When an incident Weyl fermion is totally reflected at the $z=0$ plane, $J_z$ is conserved, i.e. $J_z^I=J_z^R$, where $I$ and $R$ label incident and reflected, respectively. Since the incidence is in the $xz$ plane, $k_y=0$, $k_x$ is conserved, and $k_z$ changes sign after the reflection. Then the IF shift is
\begin{eqnarray}
\Delta_{IF}=y^R-y^I=\frac{N}{2k_x}(\langle\si_z^R\rangle-\langle\si_z^I\rangle)=-\frac{Nk_z}{E k_x},
\end{eqnarray}
which agrees with the result calculated using quantum mechanics and semiclassical EOM. However, if the monopole charges are different at the two sides, it is not obvious that one can use this argument to calculate the IF shift. 

\begin{figure}
  \centering
  \subfigure[]{\includegraphics[width=4cm]{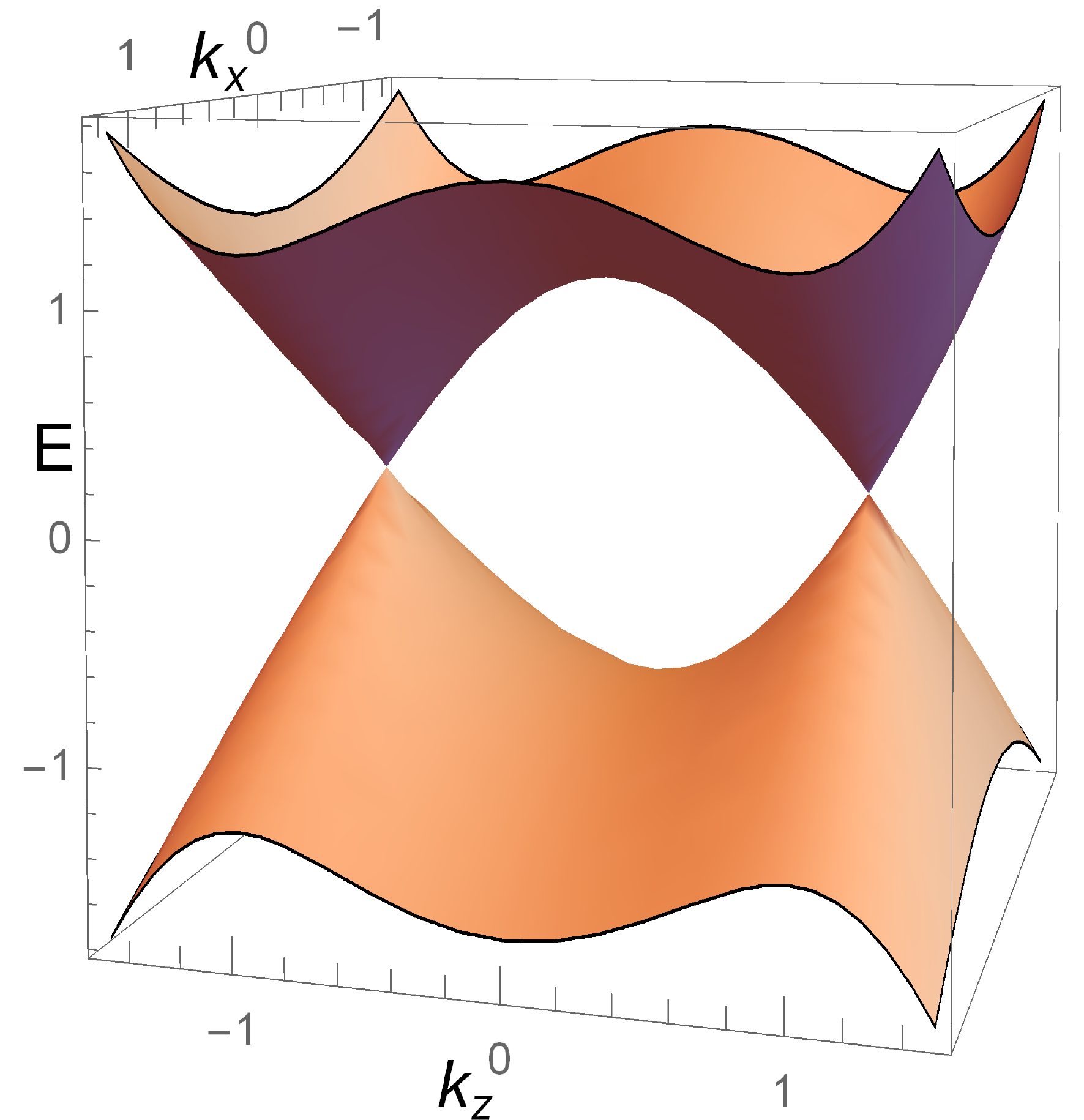}\label{fig:twocones}}
  \subfigure[]{\includegraphics[width=4cm]{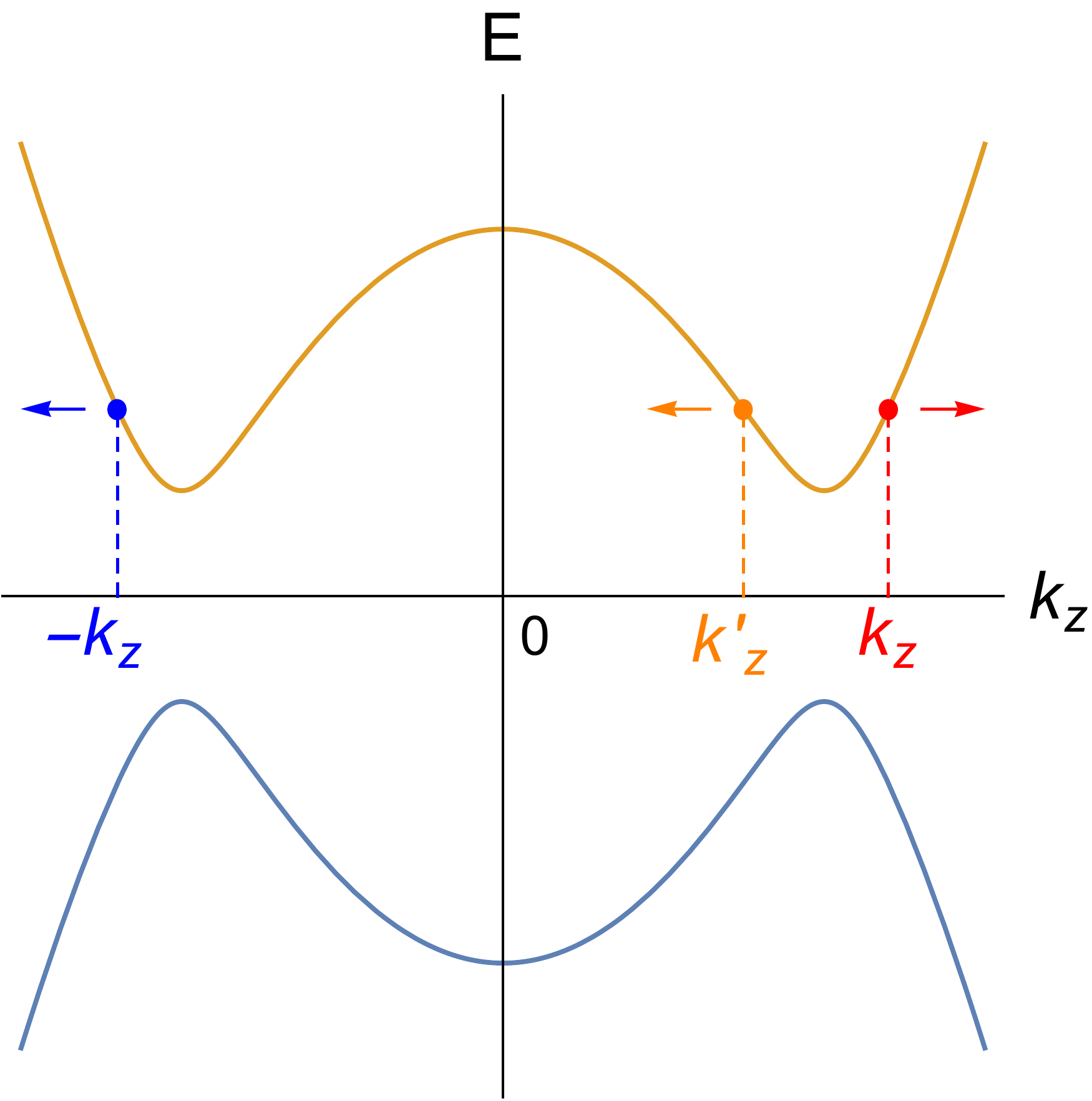}\label{fig:inter}}
  \caption{(a) Band structure of the Hamiltonian. (b) An incident particle (with fixed $k_x$) can be reflected to the same Weyl cone or the other. The red dot indicates an incident particle with momentum $k_{z}$, the orange dot indicates the case where it is reflected to the same cone with momentum $k_{z}'$, and the blue dot indicates the case where it is reflected to the other cone with momentum $-k_{z}$.}
\end{figure}
\begin{figure}
  \centering
  \subfigure[]{\includegraphics[width=4cm]{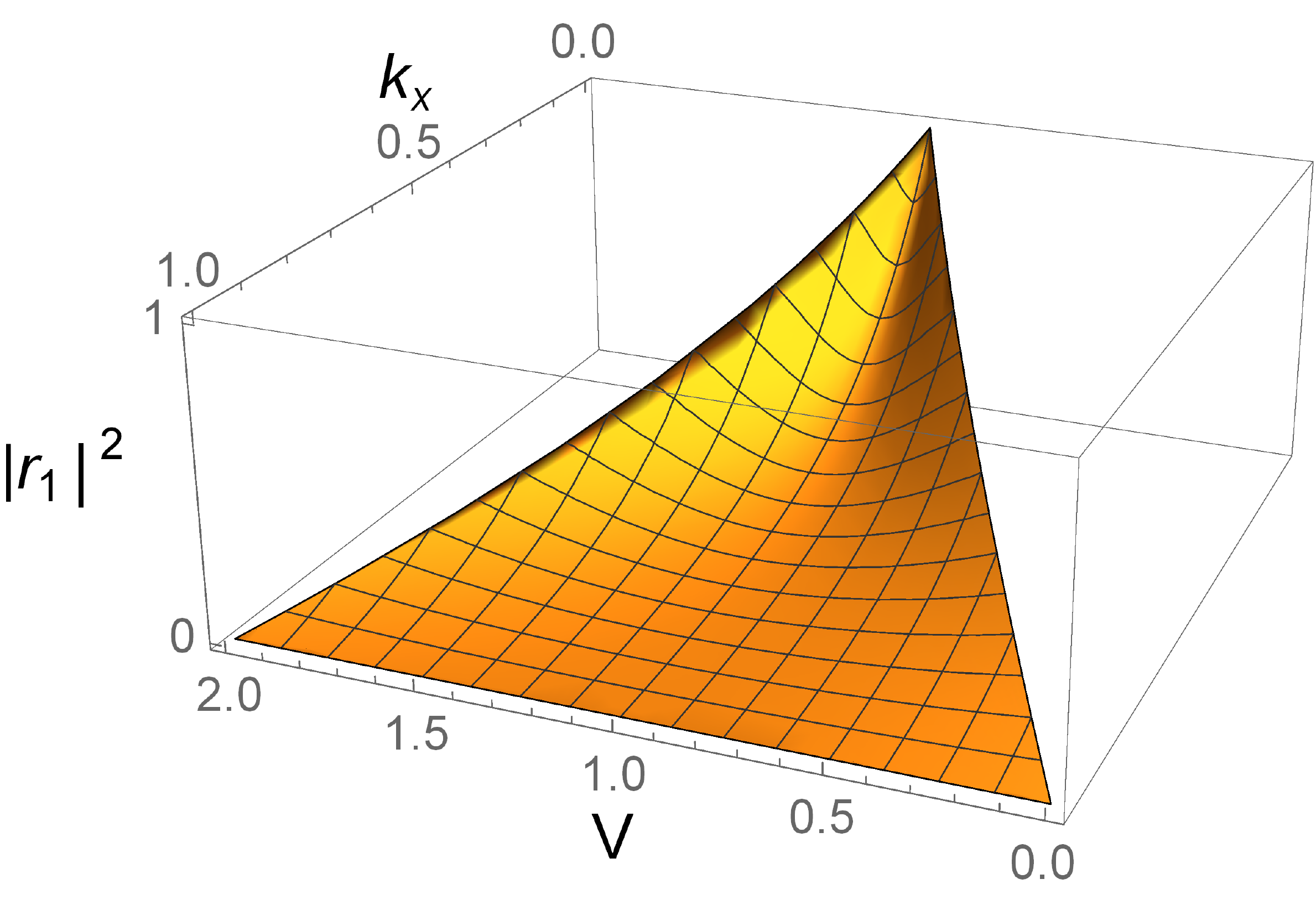}\label{fig:r11}}
  \subfigure[]{\includegraphics[width=4cm]{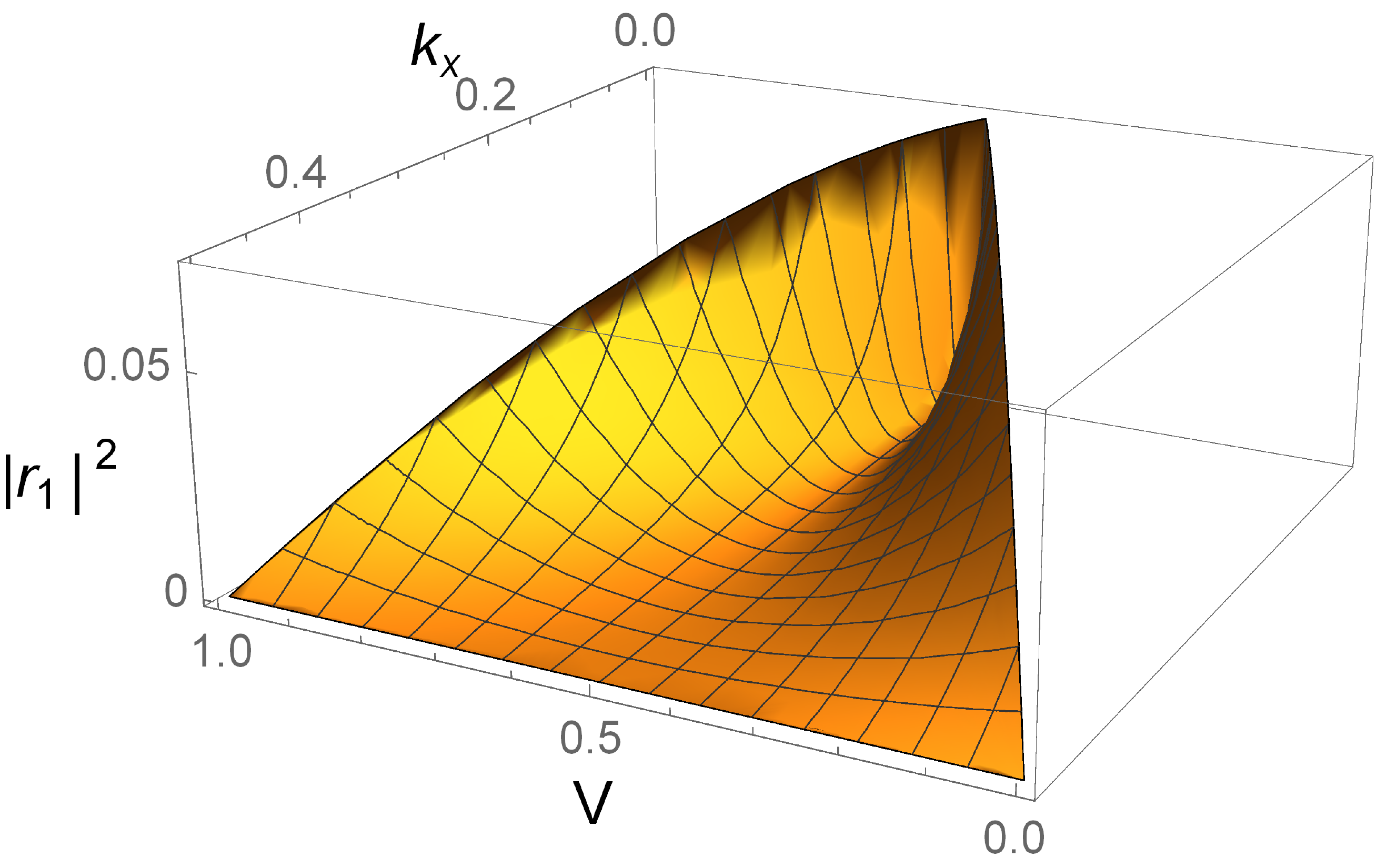}\label{fig:r12}}
  \caption{Intervalley reflectance $|r_1|^2$ with the energy of the incident Weyl fermions at $E=k_0^2$ (a) and at $E=0.5k_0^2$ (b), where we set $k_0=1$. 
  }\label{fig:r}
\end{figure}

\section{Intervalley scattering}
Previous discussions of the IF shift in WSMs\cite{xie-2015,yang-2015b} have been restricted to the case where the reflected Weyl fermions are in the same valley as the incident ones. However, since WSMs always hold an even number of Weyl points\cite{nielsen-1983}, there is a certain probability that in total reflection, part of the beam is reflected to another valley. We study this case using a model Hamiltonian with two Weyl cones of monopole charge $N=\pm 1$ located at $(0,0,\pm k_0)$, where $k_0$ is a constant,
\begin{eqnarray}
  H(\bk) &=& \left(\begin{array}{cc}
                k_z^2-k_0^2 & k_x-ik_y\\
                k_x+ik_y & -(k_z^2-k_0^2)
              \end{array}\right).
\end{eqnarray}
The dispersion is shown in Fig.\ref{fig:twocones}, and the eigenfunction is
\begin{eqnarray}
  \psi(\br) &=& \frac{1}{\sqrt{1+\eta^2}}\left(\begin{array}{c}
                    e^{-i\al} \\
                    \eta
                  \end{array}\right)e^{i(k_xx+k_yy+k_zz)}
\end{eqnarray}
where $k_z^2=\sqrt{E^2-k_x^2-k_y^2}+k_0^2$, $\eta=\sqrt{\frac{E-k_z^2+k_0^2}{E+k_z^2-k_0^2}}$ and $\al$ is defined the same way as above.

The setting is the same as in Fig.\ref{fig:refl} except that two valleys are considered. As shown in Fig.\ref{fig:inter}, the incident wave has $z$-momentum $k_z$ (red point), and by energy conservation, the $z$-momentum of the reflected wave is either $-k_z$ (blue point) or $k_z'=\sqrt{2k_0^2-k_z^2}$ (orange point), corresponding to the intervalley and intravalley scattering. 
We have the reflected wavefunction in the region $z<0$ (ignoring the factor $e^{i(k_xx+k_yy)}$)
\begin{eqnarray}
 \psi_R({\bf r})= \frac{r_1 e^{-ik_zz}}{\sqrt{1+\eta^2}}\left(\begin{array}{c}
                    e^{-i\al} \\
                    \eta
                  \end{array}\right)
  +\frac{r_2 e^{ik_z'z}}{\sqrt{1+\eta^2}}\left(\begin{array}{c}
                    \eta e^{-i\al} \\
                    1
                  \end{array}\right),
\end{eqnarray}
where $r_1$ and $r_2$ are the reflective coefficients associated with the intervalley and intravalley scattering, respectively.

Using the continuity of the wavefunction and of the derivative of the wave function, we can find $r_1$ and $r_2$ (see Appendix B for details). $|r_1|^2$ is the probability that the particle is reflected to the other valley, and $|r_2|^2|v_2|/v_1$ is the probability that it is reflected to the same valley, where $v_1>0$ is the $z$-component of the incident velocity and $-v_1$ and $v_2$ are the $z$-component of the velocities of particles that are reflected to the other and the same valley, respectively. The latter has a factor of $|v_2|/v_1$ because of the conservation of the probability current $v_1=|r_1|^2v_1+|r_2|^2|v_2|$.

We plot $|r_1|^2$ for two energies, $E=k_0^2$ and $E=0.5k_0^2$ in Fig.\ref{fig:r}. The parameters $0<V_0<2E$ and $|E-V_0|<k_x<E$ so that total reflection occurs. $|r_1|^2$ strongly depends on the energy of the particles, since the potential barrier is shallower for higher energy. We see that at smaller incident angles (smaller $k_x$), there is a larger probability of reflecting to the other valley; as $V_0\to E$, $|r_1|^2$ reaches the maximum at $k_x\to 0$. The maximum can be nearly $100\%$ when the energy of the incident particles is high, but decreases to a few percent if the energy decreases to its half. Therefore, the intervalley scattering is weak if the energy is well below the potential barrier, but in general the beam splits into two.

To calculate the shift, we find the initial center of beam is $\partial_{k_y}\al/(1+\eta^2)|_{k_y=0}$, while the final center for the two reflected beams are, respectively, $[\partial_{k_y}\al/(1+\eta^2)-\partial_{k_y}\phi_{r_1}]|_{k_y=0}$ and $[\eta^2\partial_{k_y}\al/(1+\eta^2)-\partial_{k_y}\phi_{r_2}]|_{k_y=0}$. So the shift for the two beams corresponding to intervalley and intravalley reflection are given by
\begin{eqnarray}
  \Delta_1 &=& -\partial_{k_y}\phi_{r_1}|_{k_y=0},\\
  \Delta_2 &=& -\partial_{k_y}\phi_{r_2}|_{k_y=0} +\frac{（\eta^2-1）}{\eta^2+1}\partial_{k_y}\al|_{k_y=0},
\end{eqnarray}
respectively. Since $k_y$ always appears as $k_y^2$ except in $\al$, we have $\partial_{k_y}\phi_{r_{1,2}}|_{k_y=0}=0$. Therefore $\Delta_1=0$, i.e., the beam reflected to the other valley has no IF shift, and the intravalley reflected beam has the IF shift $\Delta_2=-v_2(k_z-k_0)/(Ek_x)$. Note the replacement of $k_z$ by $k_z-k_0$ due to the finite momentum location of the Weyl point.

The zero IF shift can be easily understood from the semiclassical point of view. If the two Weyl points with monopole charge $\pm N$ are located at $(0,0,\pm k_{0})$, as the $z$-momentum goes from $k_z$ to $-k_z$, the trajectory in $k$-space passes both $k_{0}$ and $-k_{0}$ or none of the two. Then we have
\begin{eqnarray}
\Delta_{IF}&=&\int_{k_z}^{-k_z}dk_z\Omega_x=\int_{k_z}^{-k_z}dk_z(\Omega_x^{(1)}+\Omega_x^{(2)})=0,
\end{eqnarray}
where $\Omega_x^{(1,2)}$ are the Berry curvature contributed by the two Weyl points. Actually, the vanishing of the IF shift is due to the inversion symmetry and the rotational symmetry about the $z$-axis: the inversion symmetry gives ${\bf\Omega}(-\bk)={\bf\Omega}(\bk)$, from which $\Omega_x(-k_x,0,-k_z)=\Omega_x(k_x,0,k_z)$, while from the rotational symmetry, $\Omega_x(-k_x,0,k_z)=-\Omega_x(k_x,0,k_z)$. Combining the two symmetries, we have $\Omega_x(k_x,0,-k_z)=-\Omega_x(k_x,0,k_z)$, which says $\Omega_x$ is an odd function of $k_z$. Thus the integral vanishes, i.e., $\Delta_{IF}=0$.

\section{Discussion}
The IF shift in WSMs could be observed in the confirmed WSMs TaAs and NbP\cite{lv-2015a,xu-2015a,yang-2015a,lv-2015b,xu-2015b}. The dependence of the IF shift on intervalley scattering should be easily detected since there are many Weyl points in these materials. More specific calculations need to be carried out, following the same principles given here. The possible materials that host multi-Weyl points include HgCr$_2$Se$_4$\cite{xu-2011} and SrSi$_2$\cite{huang-2015b}, which provide platforms to observe the monopole charge dependence of the IF shift.

The discovery of WSMs inspires the realization of (multi-)Weyl points in systems besides solid state materials, such as photonic crystals\cite{lu-2013,lu-2015,chen-2016}. Around Weyl points, the states are governed by the same Weyl equation, so we expect that our formalism and results also apply to these systems. However, our two-Weyl cones model breaks time-reversal symmetry, and the two Weyl points are related by mirror symmetry. In a time-reversal invariant photonic Weyl crystal, a minimal number of four Weyl points can be realized\cite{lu-2015,wang-2015}. If an interface exists perpendicular to the axis at which two Weyl points related by time-reversal symmetry reside, intervalley scattering occurs between the two valleys which have the same chirality. In this case, the IF shift contributed by the two valleys are the same rather than opposite, so the shift is approximately twice that of the intravalley reflection.  Consequently, our results may be used to detect the topology of various kinds of materials by experimentally measuring their IF shift.

\begin{acknowledgements}
We would like to thank Hong Yao for helpful discussions. This work was supported in part by the NSFC under Grant No. 11474175 at Tsinghua University (LW and SKJ).
\end{acknowledgements}

\appendix
\begin{widetext}
\section{Quantum mechanical approach}
In this appendix we solve the IF shift quantum mechanically. The Hamiltonian is
\begin{eqnarray}
H&=&\left(\begin{array}{cc}k_z&(k_x-ik_y)^{N_1}\\(k_x+ik_y)^{N_1}&-k_z\end{array}\right) =\left(\begin{array}{cc}k_z&k_\parallel^{N_1}e^{-iN_1\al}\\k_\parallel^{N_1}e^{iN_1\al}&-k_z\end{array}\right) \mbox{ for }z<0,\\
H&=&\left(\begin{array}{cc}k_z&(k_x-ik_y)^{N_2}\\(k_x+ik_y)^{N_2}&-k_z\end{array}\right)+V_0 =\left(\begin{array}{cc}k_z&k_\parallel^{N_2}e^{-iN_2\al}\\k_\parallel^{N_2}e^{iN_2\al}&-k_z\end{array}\right)+V_0 \mbox{ for }z>0,
\end{eqnarray}
where $\al=\tan^{-1}\frac{k_y}{k_x}$ and $\eta=\sqrt{\frac{E-k_z}{E+k_z}}$. $E=\sqrt{k_\parallel^{2N_1}+k_z^2}$. During the process of reflection, $k_x$ and $k_y$ are conserved, while $k_z$ is not. The wavefunction at $z<0$ consists of the incident and reflected wavefunction, so
\begin{eqnarray}
  \psi_1(\br) &=& \frac{1}{\sqrt{1+\eta^2}}e^{i(k_xx+k_yy+k_zz)}\left(\begin{array}{c}
                                                                        e^{-iN_1\al} \\
                                                                        \eta
                                                                      \end{array}\right)
              +\frac{r}{\sqrt{1+\eta^2}}e^{i(k_xx+k_yy-k_zz)}\left(\begin{array}{c}
                                                                        \eta e^{-iN_1\al}\\
                                                                        1
                                                                      \end{array}\right) \mbox{ for }z<0,\\
  \psi_2(\br) &=& \frac{t}{\sqrt{1+|\xi|^2}}e^{i(k_xx+k_yy)-\kappa z}\left(\begin{array}{c}
                                                                        \xi e^{-iN_2\al} \\
                                                                        1
                                                                      \end{array}\right)
               = \frac{t}{\sqrt{2}}e^{i(k_xx+k_yy)-\kappa z}\left(\begin{array}{c}
                                                                        e^{-i\phi_\xi-iN_2\al}\\
                                                                        1
                                                                      \end{array}\right) \mbox{ for }z>0,
\end{eqnarray}
where $\kappa=\sqrt{k_\parallel^{2N_2}-(E-V_0)^2}$, $\phi_\xi=-\tan^{-1}\frac{\kappa}{E-V_0}$ if $E>V_0$, and $\phi_\xi=-\tan^{-1}\frac{\kappa}{E-V_0}+\pi$ if $E<V_0$. We have used the relation $\xi=\frac{k_\parallel^{N_2}}{E-V_0-i\kappa}=\frac{k_\parallel^{N_2}}{\sqrt{(E-V_0)^2+\kappa^2}e^{i\phi_\xi}}=e^{-i\phi_\xi}$. Connecting the two wave functions at $z=0$, we have
\begin{eqnarray}
   \frac{(1+r\eta)e^{-iN_1\al}}{\sqrt{1+\eta^2}}&=& \frac{t e^{-i\phi_\xi-iN_2\al}}{\sqrt{2}}, \\
   \frac{\eta+r}{\sqrt{1+\eta^2}} &=& \frac{t}{\sqrt{2}}.
\end{eqnarray}
The ratio of the two equations yields
\begin{eqnarray}
  \frac{1+r\eta}{\eta+r} &=& e^{-i\phi_\xi+i(N_1-N_2)\al}.
\end{eqnarray}
Solving for $r$, we get
\begin{eqnarray}
  r &=& \frac{1-\eta e^{-i\phi_\xi+i(N_1-N_2)\al}}{e^{-i\phi_\xi+i(N_1-N_2)\al}-\eta}=e^{i\phi_\xi-i(N_1-N_2)\al}\frac{1-\eta e^{-i\phi_\xi+i(N_1-N_2)\al}}{1-\eta e^{i\phi_\xi-i(N_1-N_2)\al}}\equiv e^{i\phi_r},
\end{eqnarray}
so the phase shift is
\begin{eqnarray}
  \phi_r &=& \phi_\xi-(N_1-N_2)\al+2\tan^{-1}\frac{\eta\sin(\phi_\xi-(N_1-N_2)\al)}{1-\eta\cos(\phi_\xi-(N_1-N_2)\al)}.
\end{eqnarray}
$\phi_\xi$ is even in $k_y$, so $\partial_{k_y}\phi_\xi|_{k_y=0}=0$. Then
\begin{eqnarray}
  \partial_{k_y}\phi_r|_{k_y=0} &=& \frac{(N_1-N_2)k_z}{(E-V_0)k_x^{N_1-N_2+1}-Ek_x}.
\end{eqnarray}
Assuming the beams are Gaussian. The center of the incident beam is $\frac{N_1\partial_{k_y}\al|_0}{1+\eta^2}$, and the center of the reflected beam is
\begin{eqnarray}
N_1\frac{(\partial_{k_y}\al-\partial_{k_y}\phi_r)|_0\eta^2-\partial_{k_y}\phi_r}{1+\eta^2}=\frac{N_1\eta^2\partial_{k_y}\al|_0}{1+\eta^2}-\partial_{k_y}\phi_r.
\end{eqnarray}
So the IF shift is
\begin{eqnarray}
  \Delta_{IF} &=& N_1\partial_{k_y}\al|_0\frac{\eta^2-1}{\eta^2+1}-\partial_{k_y}\phi_r|_0 =-\frac{N_1k_z}{Ek_x}+\frac{(N_1-N_2)k_z}{Ek_x-(E-V_0)k_x^{N_1-N_2+1}}.
\end{eqnarray}
\section{Probability of intervalley scattering}
There is a certain probability that on total reflection, the particle is reflected to the other Weyl cone. Now we calculate this probability using a quantum mechanical approach. We start with the Hamiltonian
\begin{eqnarray}
  H &=& \left(\begin{array}{cc}
                k_z^2-k_0^2 & k_x-ik_y\\
                k_x+ik_y & -(k_z^2-k_0^2)
              \end{array}\right)
\end{eqnarray}
where $k_i$'s are operators and $k_i=-i\partial_{x_i}$. The eigenfunction is
\begin{eqnarray}
  \psi(x,y,z) &=& \frac{1}{\sqrt{1+\eta^2}}\left(\begin{array}{c}
                    e^{-i\al} \\
                    \eta
                  \end{array}\right)e^{i(k_xx+k_yy+k_zz)}
\end{eqnarray}
where $k_z=\sqrt{\sqrt{E^2-k_x^2-k_y^2}+k_0^2}$, $\eta=\frac{E-(k_z^2-k_0^2)}{k_\parallel}=\sqrt{\frac{E-(k_z^2-k_0^2)}{E+(k_z^2-k_0^2)}}$ and $\al=\tan^{-1}\frac{k_y}{k_x}$. Assume there is an interface at $z=0$, and a potential
\begin{equation}
  V=\left\{\begin{array}{cc}
             0,&z<0 \\
             V_0.&z>0
           \end{array}\right.
\end{equation}
Upon reflection, $k_x$ and $k_y$ are conserved due to translational symmetry, while $k_z$ is not. By energy conservation, the $z$-momentum of the reflected particle is either $-k_z$ or $k_z'=\sqrt{2k_0^2-k_z^2}$. (We assume $k_z>0$.) The former is the case in which the particle is reflected to the other Weyl cone. Note that with $k_z\to -k_z, \eta\to\eta$; while with $k_z\to k_z', \eta\to 1/\eta$. Ignoring the factor $e^{i(k_xx+k_yy)}$, we have the wavefunction in the region $z<0$ and $z>0$
\begin{eqnarray}
  \psi_1(z)&=& \frac{1}{\sqrt{1+\eta^2}}\left(\begin{array}{c}
                    e^{-i\al} \\
                    \eta
                  \end{array}\right)e^{ik_zz}
  +\frac{r_1}{\sqrt{1+\eta^2}}\left(\begin{array}{c}
                    e^{-i\al} \\
                    \eta
                  \end{array}\right)e^{-ik_zz}
  +\frac{r_2}{\sqrt{1+\eta^2}}\left(\begin{array}{c}
                    \eta e^{-i\al} \\
                    1
                  \end{array}\right)e^{ik_z'z},\\
  \psi_2(z)&=& \frac{t_1}{\sqrt{1+|\xi_1|^2}}\left(\begin{array}{c}
                    \xi_1e^{-i\al} \\
                    1
                  \end{array}\right)e^{ik_1z}
  +\frac{t_2}{\sqrt{1+|\xi_2|^2}}\left(\begin{array}{c}
                    \xi_2 e^{-i\al} \\
                    1
                  \end{array}\right)e^{ik_2z}\\
  &=& \frac{t_1}{\sqrt{2}}\left(\begin{array}{c}
                    e^{i\phi_\xi}e^{-i\al} \\
                    1
                  \end{array}\right)e^{ik_1z}
  +\frac{t_2}{\sqrt{2}}\left(\begin{array}{c}
                    e^{-i\phi_\xi} e^{-i\al} \\
                    1
                  \end{array}\right)e^{ik_2z}.
\end{eqnarray}
If $(E-V_0)^2<k_x^2+k_y^2$, to satisfy $E=\sqrt{k_x^2+k_y^2+(k_{1,2}^2-k_0^2)}+V_0$, we must have $k_{1,2}^2-k_0^2=\pm i\kappa$ where $\kappa=\sqrt{k_x^2+k_y^2-(E-V_0)^2}>0$. The imaginary part of $k_1$ and $k_2$ should be positive, so that the wave function decays at $z>0$ side, otherwise the wave function would explode. Thus we have $k_1=(k_0^4+\kappa^2)^\frac{1}{4}e^{i\phi_k}$ and $k_2=-(k_0^4+\kappa^2)^\frac{1}{4}e^{-i\phi_k}$ where $\phi_k=\frac{1}{2}\tan^{-1}\frac{\kappa}{k_0^2}$. Then $\xi_1=\frac{k_\parallel}{E-V_0-(k_1^2-k_0^2)}=\frac{k_\parallel}{E-V_0-i\kappa}=e^{i\phi_\xi}$, and $\xi_2=\frac{k_\parallel}{E-V_0+i\kappa}=e^{-i\phi_\xi}$ where $\phi_\xi=\tan^{-1}\frac{\kappa}{E-V_0}$ if $E-V_0>0$ and $\phi_\xi=\tan^{-1}\frac{\kappa}{E-V_0}-\pi$ if $E-V_0<0$. The boundary conditions at $z=0$ are $\psi_1(0)=\psi_2(0)$ and $\psi_1'(0)=\psi_2'(0)$, explicitly,
\begin{eqnarray}
  \frac{1+r_1+r_2\eta}{\sqrt{1+\eta^2}} &=& \frac{t_1e^{i\phi_\xi}+t_2e^{-i\phi_\xi}}{\sqrt{2}},\\
  \frac{(1+r_1)\eta+r_2}{\sqrt{1+\eta^2}} &=& \frac{t_1+t_2}{\sqrt{2}},\\
  \frac{(1-r_1)k_z+r_2\eta k_{z1}}{\sqrt{1+\eta^2}} &=& \frac{(k_0^4+\kappa^2)^{\frac{1}{4}}(t_1e^{i(\phi_k+\phi_\xi)}-t_2e^{-i(\phi_k+\phi_\xi)})}{\sqrt{2}},\\
  \frac{(1-r_1)\eta k_z+r_2 k_{z1}}{\sqrt{1+\eta^2}} &=& \frac{(k_0^4+\kappa^2)^{\frac{1}{4}}(t_1e^{i\phi_k}-t_2e^{-i\phi_k})}{\sqrt{2}}.
\end{eqnarray}
The reflected beam splits into two beams. We solve the reflectance $|r_1|^2$ and $|r_2|^2$ and verify the conservation of the current $v_i=|r_1|^2v_1+|r_2|^2v_2$ numerically, where $v_i$ is the incident velocity and $v_1$ and $v_2$ are the velocities of the two reflected beams ($v_1=v_i$).

\end{widetext}


\begin{thebibliography}{99}

\bibitem{sinova-2015} J. Sinova, S. O. Valenzuela, J. Wunderlich, C. H. Back, and T. Jungwirth, Rev. Mod. Phys. {\bf 87}, 1213 (2015).

\bibitem{maciejko-2011} J. Maciejko, T. L. Hughes, and S.-C. Zhang, Annu. Rev. Condens. Matter Phys. {\bf 2}, 31 (2011).

\bibitem{hasan-2010} M. Z. Hasan, and C. L. Kane, Rev. Mod. Phys. {\bf 82}, 3045 (2010).

\bibitem{qi-2011} X.-L. Qi, and S.-C. Zhang, Rev. Mod. Phys. {\bf 83}, 1057 (2011).

\bibitem{bliokh-2015} K. Y. Bliokh, F. J. Rodr\'iguez-Fortu\~no, F. Nori, and A. V. Zayats, Nat. Phot. {\bf 9}, 796 (2015).

\bibitem{onoda-2004} M. Onoda, S. Murakami, and N. Nagaosa, Phys. Rev. Lett. {\bf 93}, 083901 (2004).

\bibitem{bliokh-2006} K. Y. Bliokh, and Y. P. Bliokh, Phys. Rev. Lett. {\bf 96}, 073903 (2006).

\bibitem{fedorov-1955} F. Fedorov, Dokl. Akad. Nauk SSSR {\bf 105}, 465 (1955).

\bibitem{imbert-1972} C. Imbert, Phys. Rev. D {\bf 5}, 787 (1972).

\bibitem{hosten-2008} O. Hosten and P. Kwiat, Science {\bf 319}, 787 (2008).

\bibitem{yin-2013} X. Yin, Z. Ye, J. Rho, Y. Wang, and X. Zhang, Science {\bf 339}, 1405 (2013).

\bibitem{xie-2015} Q.-D. Jiang, H. Jiang, H. Liu, Q.-F. Sun, and X. C. Xie, Phys. Rev. Lett. {\bf 115}, 156602 (2015).

\bibitem{yang-2015b} S. A. Yang, H. Pan, and F. Zhang, Phys. Rev. Lett. {\bf 115}, 156603 (2015).

\bibitem{lv-2015a} B. Q. Lv, H. M. Weng, B. B. Fu, X. P. Wang, H. Miao, J. Ma, P. Richard, X. C. Huang, L. X. Zhao, G. F. Chen, Z. Fang, X. Dai, T. Qian, and H. Ding, Phys. Rev. X {\bf 5}, 031013 (2015).

\bibitem{xu-2015a} S.-Y. Xu, I. Belopolski, N. Alidoust, M. Neupane, G. Bian, C. Zhang, R. Sankar, G. Chang, Z. Yuan, C.-C. Lee, S.-M. Huang, H. Zheng, J. Ma, D. S. Sanchez, B. Wang, A. Bansil, F. Chou, P. P. Shibayev, H. Lin, S. Jia, and M. Z. Hasan, Science {\bf 349}, 613 (2015).

\bibitem{yang-2015a} L. X. Yang, Z. K. Liu, Y. Sun, H. Peng, H. F. Yang, T. Zhang, B. Zhou, Y. Zhang, Y. F. Guo, M. Rahn, D. Prabhakaran, Z. Hussain, S. K. Mo, C. Felser, B. Yan, and Y. L. Chen, Nature Physics {\bf 11}, 728 (2015).

\bibitem{lv-2015b}B. Q. Lv, N. Xu, H. M. Weng, J. Z. Ma, P. Richard, X. C. Huang, L. X. Zhao, G. F. Chen, C. E. Matt, F. Bisti, V. N. Strocov, J. Mesot, Z. Fang, X. Dai, T. Qian, M. Shi, and H. Ding, Nature Physics {\bf11}, 724 (2015).

\bibitem{xu-2015b} S.-Y. Xu, N. Alidoust, I. Belopolski, Z. Yuan, G. Bian, T.-R. Chang, H. Zheng, V. N. Strocov, D. S. Sanchez, G. Chang, C. Zhang, D. Mou, Y. Wu, L. Huang, C.-C. Lee, S.-M. Huang, B. Wang, A. Bansil, H.-T. Jeng, T. Ne- upert, A. Kaminski, H. Lin, S. Jia, and M. Zahid Hasan, Nature Physics {\bf11}, 748 (2015).

\bibitem{weyl-1929} H. Weyl, Z. Phys. {\bf 56}, 330 (1929).

\bibitem{volovik-2009} G. E. Volovik, {\it The universe in a helium droplet}, Oxford University Press (2009).

\bibitem{wan-2011} X. Wan, A. M. Turner, A. Vishwanath, and S. Y. Savrasov, Phys. Rev. B {\bf 83}, 205101 (2011).

\bibitem{xu-2011} G. Xu, H. Weng, Z. Wang, X. Dai, and Z. Fang, Phys. Rev. Lett. {\bf 107}, 186806 (2011).

\bibitem{burkov-2011} A. A. Burkov and L. Balents, Phys. Rev. Lett. {\bf 107}, 127205 (2011).

\bibitem{yang-2011} K.-Y. Yang, Y.-M. Lu, and Y. Ran, Phys. Rev. B {\bf 84}, 075129 (2011).

\bibitem{halasz-2012} G. B. Halasz and L. Balents, Phys. Rev. B {\bf 85}, 035103 (2012).

\bibitem{zhang-2014} H. Zhang, J. Wang, G. Xu, Y. Xu, and S.-C. Zhang, Phys. Rev. Lett. {\bf 112}, 096804 (2014).

\bibitem{liu-2014} J. Liu and D. Vanderbilt, Phys. Rev. B {\bf 90}, 155316 (2014).

\bibitem{weng-2015} H. Weng, C. Fang, Z. Fang, B. A. Bernevig, and X. Dai, Phys. Rev. X {\bf 5}, 011029 (2015).

\bibitem{huang-2015} S.-M. Huang, S.-Y. Xu, I. Belopolski, C.-C. Lee, G. Chang, B. Wang, N. Alidoust, G. Bian, M. Neupane, C. Zhang, S. Jia, A. Bansil, H. Lin, and M. Z. Hasan, Nature Communications {\bf 6}, 7373 (2015).

\bibitem{hirayama-2015} M. Hirayama, R. Okugawa, S. Ishibashi, S. Murakami, and T. Miyake, Phys. Rev. Lett. {\bf 114}, 206401 (2015).

\bibitem{xiao-2010} D. Xiao, M.-C. Chang, and Q. Niu, Rev. Mod. Phys. {\bf 82}, 1959 (2010).

\bibitem{fang-2012} C. Fang, M. J. Gilbert, X. Dai, and B. A. Bernevig, Phys. Rev. Lett. {\bf 108}, 266802 (2012).


\bibitem{zhang-2011} F. Zhang, J. Jung, G. A. Fiete, Q. Niu, and A. H. MacDonald, Phys. Rev. Lett. {\bf 106}, 156801 (2011).

\bibitem{thouless-1982} D. J. Thouless, M. Kohmoto, M. P. Nightingale, and M. den Nijs, Phys. Rev. Lett. {\bf 49}, 405 (1982).

\bibitem{onoda-2006} M. Onoda, S. Murakami, and N. Nagaosa, Phys. Rev. E {\bf 74}, 066610 (2006).

\bibitem{nielsen-1983} H. B. Nielsen, and M. Ninomiya, Phys. Lett. B {\bf 130}, 389 (1983).

\bibitem{huang-2015b} S.-M. Huang, S.-Y. Xu, I. Belopolski, C.-C. Lee, G. Chang, T. R. Chang, B. Wang, N. Alidoust, G. Bian, M. Neupane, D. Sanchez, H. Zheng, H.-T. Jeng, A. Bansil, T. Neupert, H. Lin, and M. Z. Hasan, Proceedings of the National Acadamy of Sciences {\bf 113}, 5, 1180 (2016).

\bibitem{lu-2013} L. Lu, L. Fu, J. D. Joannopoulos, and M. Solja$\check{\text{c}}$i$\acute{\text{c}}$, Nature Photonics {\bf 7}, 294 (2013).

\bibitem{lu-2015} L. Lu, Z. Wang, D. Ye, L. Ran, L. Fu, J. D. Joannopoulos, and M. Solja$\check{\text{c}}$i$\acute{\text{c}}$, Science {\bf 349}, 622 (2015).

\bibitem{chen-2016} W.-J. Chen, M. Xiao, and C. T. Chan, Nature Communications {\bf 7}, 13038 (2016).

\bibitem{wang-2015} L. Wang, S.-K. Jian, and H. Yao, Phys. Rev. A {\bf 93}, 061801(R) (2016).

\end{thebibliography}
\end{document}